\documentclass[conference]{IEEEtran}
\IEEEoverridecommandlockouts

\usepackage{cite}
\usepackage{amsmath,amssymb,amsfonts}
\usepackage{algorithmic}
\usepackage{graphicx}
\usepackage{textcomp}
\usepackage{xcolor}
\def\BibTeX{{\rm B\kern-.05em{\sc i\kern-.025em b}\kern-.08em
    T\kern-.1667em\lower.7ex\hbox{E}\kern-.125emX}}

\usepackage{listings}
\usepackage{minted}
\usepackage{subcaption}
\usepackage{booktabs}
\usepackage{multirow}
\usepackage{makecell}
\usepackage{array}
\usepackage{caption}
\usepackage{booktabs}   %
\usepackage{tabularx}   
\usepackage{makecell}   
\usepackage{siunitx}   
\usepackage{fancyhdr}
\usepackage{url}
\usepackage{marvosym}
\newcommand{\corr}{\textsuperscript{\Letter}}
\newcommand\hl[1]{{\em \bf \em #1}}
\begin{document}

\title{TurboFuzz: FPGA Accelerated Hardware Fuzzing for Processor Agile Verification\\
\thanks{
\IEEEauthorrefmark{3}These authors contributed equally. Haoran Wu finished this work during his internship at the SKLP, Institute of Computing Technology, Chinese Academy of Sciences.
}
}

\author{
\IEEEauthorblockN{
Yang Zhong\IEEEauthorrefmark{1}\IEEEauthorrefmark{2}\IEEEauthorrefmark{3},
Haoran Wu\IEEEauthorrefmark{4}\IEEEauthorrefmark{3},
Xueqi Li\IEEEauthorrefmark{1}\IEEEauthorrefmark{2},
Sa Wang\IEEEauthorrefmark{1}\IEEEauthorrefmark{2},
David Boland\IEEEauthorrefmark{5},
Yungang Bao\IEEEauthorrefmark{1}\IEEEauthorrefmark{2},
Kan Shi\IEEEauthorrefmark{1}\IEEEauthorrefmark{2}\corr
}
\IEEEauthorblockA{\IEEEauthorrefmark{1}State Key Lab of Processors, Institute of Computing Technology, Chinese Academy of Sciences\\ 
\IEEEauthorrefmark{2}University of Chinese Academy of Sciences\\ 
\IEEEauthorrefmark{4}University of Cambridge, \IEEEauthorrefmark{5}The University of Sydney
\\
\Letter~Corresponding author: Kan Shi (shikan@ict.ac.cn)
}
}

\maketitle

\begin{abstract}
Verification is a critical process for ensuring the correctness of modern processors. The increasing complexity of processor designs and the emergence of new instruction set architectures (ISAs) like RISC-V have created demands for more agile and efficient verification methodologies, particularly regarding verification efficiency and faster coverage convergence. While simulation-based approaches now attempt to incorporate advanced software testing techniques such as fuzzing to improve coverage, they face significant limitations when applied to processor verification, notably poor performance and inadequate test case quality. Hardware-accelerated solutions using FPGA or ASIC platforms have tried to address these issues, yet they struggle with challenges including host-FPGA communication overhead, inefficient test pattern generation, and suboptimal implementation of the entire multi-step verification process.

In this paper, we present TurboFuzz, an end-to-end hardware-accelerated verification framework that implements the entire Test Generation-Simulation-Coverage Feedback loop on a single FPGA for modern processor verification. TurboFuzz enhances test quality through optimized test case (seed) control flow, efficient inter-seed scheduling, and hybrid fuzzer integration, thereby improving coverage and execution efficiency. Additionally, it employs a feedback-driven generation mechanism to accelerate coverage convergence. Experimental results show that TurboFuzz achieves up to 2.23× more coverage collection than software-based fuzzers within the same time budget, and up to 571× performance speedup when detecting real-world issues, while maintaining full visibility and debugging capabilities with moderate area overhead.
\end{abstract}


\section{Introduction}
As computer architecture and modern processor designs grow increasingly complex, functional verification has become both vital and time-intensive throughout the chip development cycle of modern processors. Recent studies show verification consumes up to 70\% of development efforts\cite{arch_sim_survey}. The emergence of new Instruction Set Architectures (ISAs) like RISC-V creates additional demand for more agile and efficient verification methods. Given the open-source nature of RISC-V, researchers are focusing on developing novel architectural modeling and functional simulation methods\cite{xiangshan} to ensure CPU cores and IPs can \textit{reliably} be used by the community. 

\begin{figure}[t]
\centering
\begin{subfigure}[b]{0.46\textwidth}
\includegraphics[width=\textwidth]{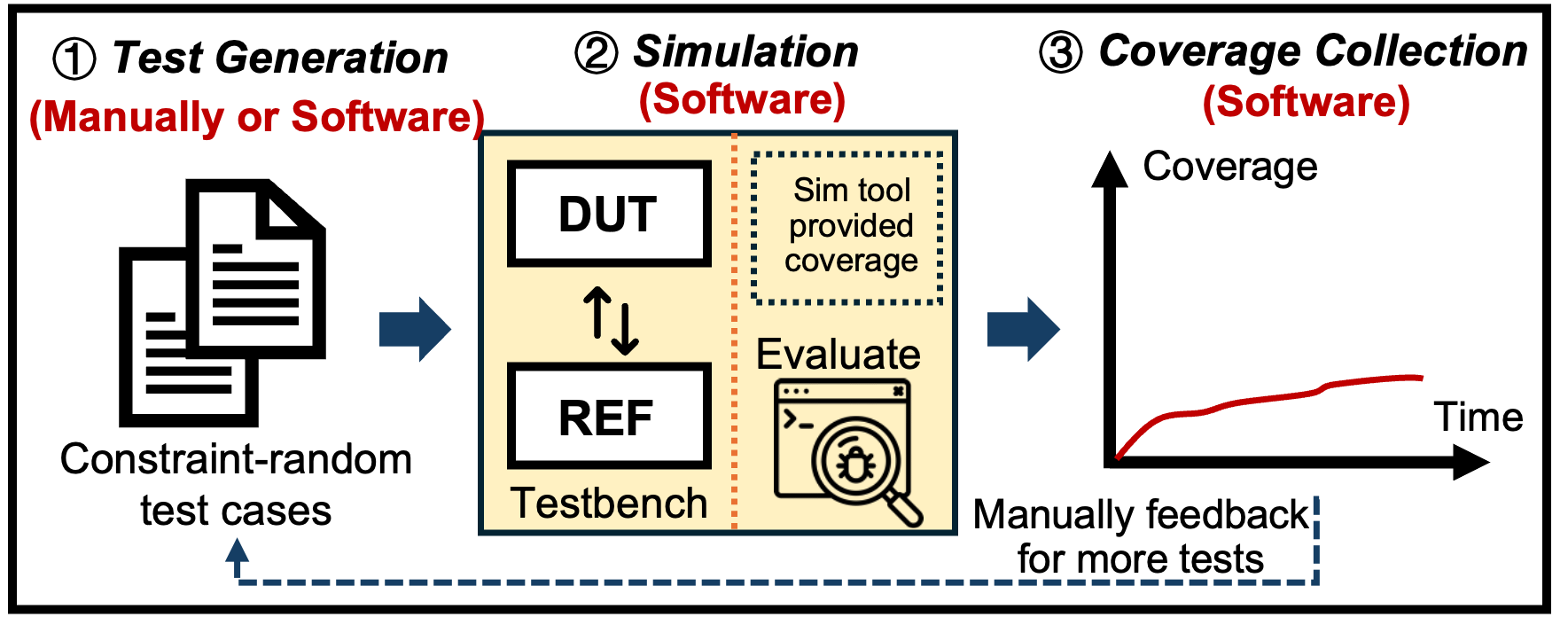}
\caption{Traditional verification flow.}
\includegraphics[width=\textwidth]{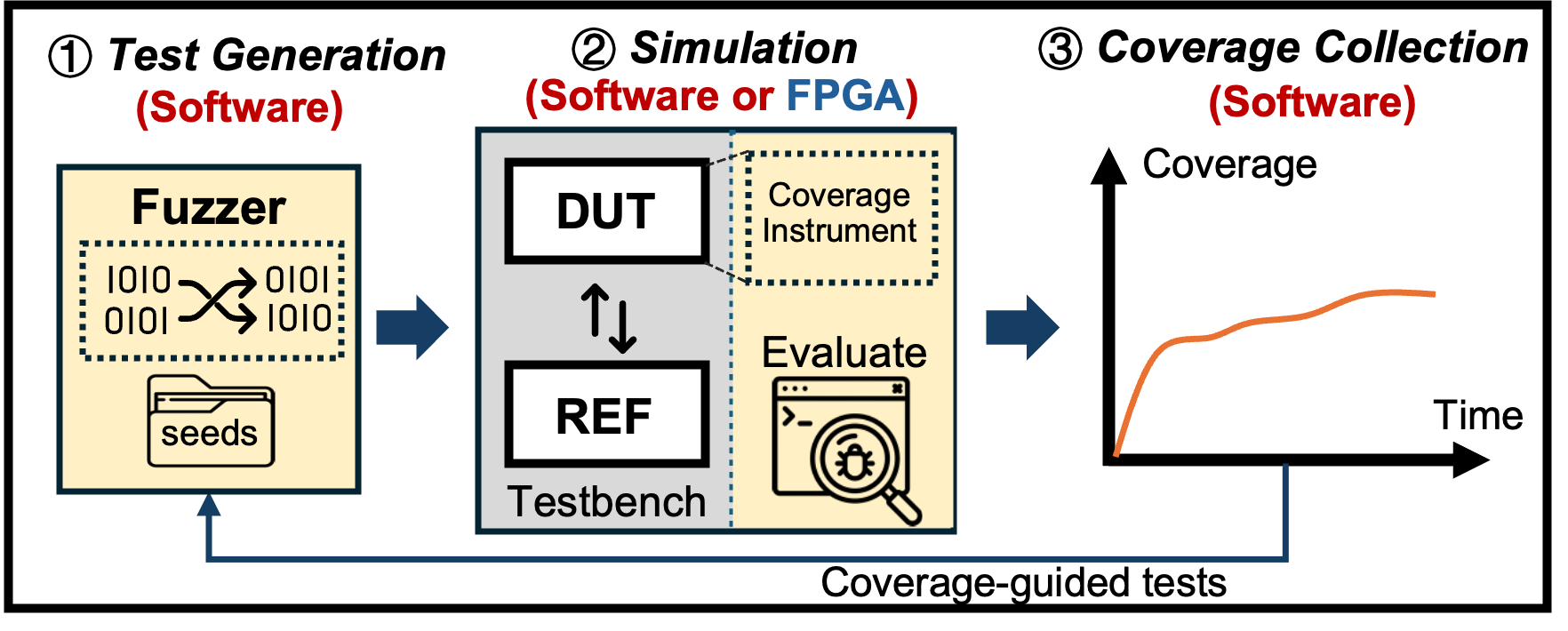}
\caption{Previous approach with software fuzzers.}
\includegraphics[width=\textwidth]{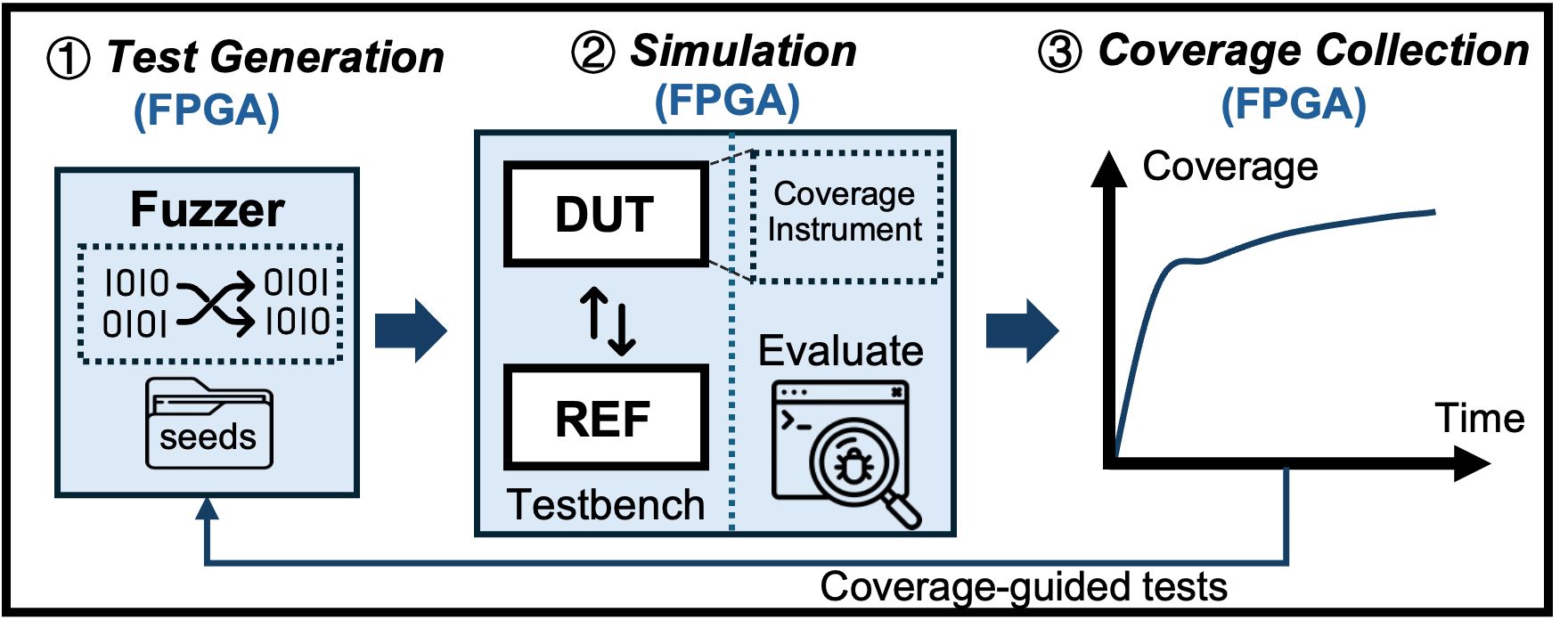}
\caption{Proposed verification flow using TurboFuzz.}
\end{subfigure}
\caption{ (a) Traditional verification flow: all steps implemented in software. (b) Advanced verification flow with software-based fuzzers: allowing coverage-guided mutation for better coverage, but still with performance bottleneck. (c) Proposed {\em TurboFuzz}: all steps implemented in FPGA for high-performance agile verification.}
 \vspace{-4ex}
\label{fig:first_page}
\end{figure}

A traditional verification flow can be characterized into three main steps, as shown in Fig.~\ref{fig:first_page} (a). 1) test generation, which creates direct or random stimulus to exercise the DUT, 2) simulation and evaluation, which feeds test vectors into both the DUT and its REF, comparing their results to identify any mismatches, and 3) coverage collection, which monitors the extent to which functional points of interest are covered during simulation. A verification engineer must create additional test cases and repeat these steps until full coverage is achieved. 

However, conventional verification faces low efficiency\cite{abv_survey_2022,Registerinformation} due to two major challenges. Firstly, conventional verification methodologies, like Universal Verification Methodology (UVM), rely heavily on software simulation. Although software simulation provides full visibility of design details for debugging, it runs extremely slowly — typically at tens of kHz. To address this, FPGA-based prototyping is widely used in industry for accelerated pre-silicon verification, as it allows RTL designs to be implemented directly on FPGAs that run at MHz speeds, enabling faster bug detection. However, debugging is challenging as FPGA-based tools such as ILA\cite{ila} and SignalTap\cite{signaltap} struggle to store operational traces and impose significant area overhead.

Another challenge lies in verification effectiveness, which strongly depends on test case quality. Traditional methods like Constrained Random Verification (CRV) inefficiently explore the space, lacking runtime control and often missing corner cases. In modern RISC-V CPUs, some corner cases can only be triggered by particular combinations of complex instructions executed in sequence. Hence, they cannot be thoroughly verified using existing CRV-based tools like RISC-V tests\cite{riscv-tests}.

Recent research has adapted software fuzzing\cite{afl-fuzz,Whitebox_fuzz,full_speed_fuzz,greybox_fuzzing} techniques for hardware verification\cite{RFUZZ,Hyperfuzzing,ProcessorFuzz,Effective_processor_verification,TheHuzz,SoCFuzzer,Symbolic_fuzz,Cascade}, as shown in Fig.~\ref{fig:first_page}(b). Unlike traditional CRV-based approaches, fuzzing uses coverage feedback to guide test generation\cite{coverageDirectedTestGeneration}. The fuzzer creates initial inputs, measures coverage, and then mutates inputs to explore new states while avoiding redundancy\cite{TheFuzzOdessy}. However, these software fuzzers run slowly, achieving just 2–4 iterations per second due to input compilation and transformation overhead\cite{ProcessorFuzz}. Previous work offloads DUT execution to FPGAs, but host-FPGA data transfer remains a bottleneck\cite{DifuzzRTL}. Moreover, existing fuzzers are not friendly to debugging: they either detect crashes without revealing root causes\cite{FuzzingHardwareasSoftware}, or rely on coarse-grained comparisons between the DUT and REF after thousands of instructions\cite{RFUZZ,DifuzzRTL}.

\begin{figure*}[tb]
  \centering
  \includegraphics[width=.95\textwidth]{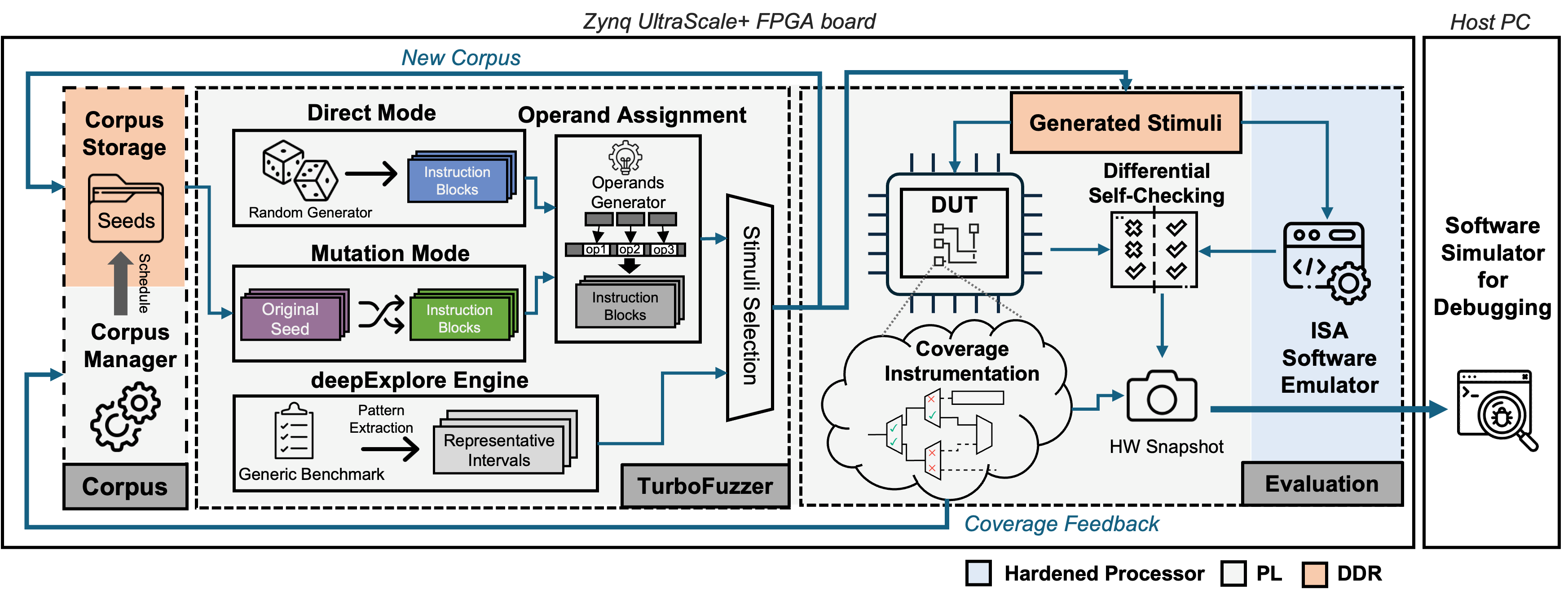}
  \caption{Overview of the TurboFuzz framework, including the TurboFuzzer, dynamic evaluation, coverage instrumentation, corpus storage and management, all implemented on a single FPGA board.}
  \label{Fig:overview}
\end{figure*}

To mitigate these bottlenecks, we propose TurboFuzz, an end-to-end verification framework with hardware acceleration, as shown in Fig.~\ref{fig:first_page}(c). TurboFuzzer incorporates all features of conventional software fuzzers while also introducing novel features, including advanced corpus scheduling for more efficient mutation, control-flow instruction optimization for better generation quality, and a hybrid approach combining direct and fuzzing methods named \textit{deepExplore} for deeper coverage space exploration. It is also fully synthesizable and directly deployed on an FPGA for fast stimulus generation.

To address the checking bottleneck, TurboFuzz implements a finer-grained self-checking scheme. We map the DUT and the REF onto the programmable logic and the hardened ARM processors of the same FPGA SoC, respectively, together with dedicated hardware monitors and checkers to track key registers and signals of both DUT and REF while maintaining instruction-level synchronization between both. This setup enables automatic, instruction-level self-checking, with both hardware and software pausing immediately on mismatches. TurboFuzz also leverages hardware snapshot technology for debugging\cite{encore_fpga23, attia2020statemover}, which could capture design states for offline analysis when mismatches occur.

Additionally, TurboFuzz optimizes the control register-based coverage instrumentation method widely employed in state-of-the-art (SOTA) fuzzing techniques\cite{DifuzzRTL,Cascade}. The existing method introduces bias towards frequently-toggled modules such as arithmetic units over control paths, and can create coverage points that cannot be reached at all. TurboFuzz addresses both issues by eliminating potential empty states, and allowing flexible coverage weighting across submodules for better design state representation. The main contributions are as follows:

\begin{itemize}
    \item We propose TurboFuzz, an end-to-end autonomous verification framework with a closed loop of hardware-accelerated operations, including test generation, hardware simulation, self-checking, and coverage collection entirely on a single FPGA SoC. 
    \item To our knowledge, TurboFuzz is the first fully synthesizable hardware fuzzer for high-performance processor verification, supporting seed mutation, improved coverage instrumentation, and deeper coverage exploration.
    
    \item TurboFuzz features a novel hybrid fuzzing approach, \textit{deepExplore}, which combines coverage-guided mutation with structural insights from real benchmarks by extracting representative instruction intervals instead of running full benchmark suites.
    
    \item Experiments on an AMD UltraScale+ FPGA with realistic processor designs. TurboFuzz achieves up to $2.23\times$ more coverage and up to $571\times$ speed-up over software approaches when catching and identifying real-world issues, with moderate area overhead. 
\end{itemize}

\section{Background} \label{sec:background}

\subsection{Hardware fuzzing}
Fuzzing originates from software testing. It continuously generates and executes randomized test inputs to discover potential vulnerabilities in the program under test. Compared to traditional directed testing, fuzzing offers greater automation and effectively explores corner cases and rare execution paths that conventional testing methods struggle to cover.

Among various fuzzing techniques, greybox fuzzing \cite{coverage_guided_greybox_fuzzing} is currently among the most widely adopted approaches. It uses runtime coverage information, such as basic block coverage, edge coverage and edge coverage, to guide the generation of subsequent test inputs. Specifically, greybox fuzzing inserts instrumentation into the target program to collect coverage information at runtime, the fuzzer selects a seed input from its corpus and generates new test inputs by applying mutation strategies. When an input triggers a new coverage point, it reveals unexplored state space. Such inputs are added to the corpus for use in subsequent fuzzing iterations.

Inspired by the success of fuzzing in software testing, researchers have increasingly applied this approach to hardware verification. RFUZZ \cite{RFUZZ} was the first to adapt the American Fuzzy Lop (AFL) framework for hardware testing. It treats software branch instructions as analogous to hardware multiplexers and uses multiplexer control signal toggling as the coverage metric. However, this metric often fails to accurately capture circuit state transitions.

DifuzzRTL \cite{DifuzzRTL} improves upon RFUZZ by adding clock-sensitive optimization and incorporating a reference model, which better captures state transitions and enables effective checking of RTL execution results. ProcessorFuzz \cite{ProcessorFuzz} introduced a coverage metric guided by processor Control and Status Register (CSR) transitions. By monitoring changes in key CSR values, ProcessorFuzz identifies when the processor enters new internal states, effectively narrowing the focus to the system's state space. MorFuzz \cite{Morfuzz} incorporates runtime instruction morphing and uses runtime feedback to generate instruction sequences that are both format-valid and semantically sound. Cascade \cite{Cascade} takes another step further by employing a program generation approach to construct longer test programs with complex control-flow and data-flow. When such a program runs on the target CPU and a bug causes a data-flow anomaly, this anomaly often triggers a chain reaction in the control flow, preventing the program from terminating as expected. Cascade promptly captures this deviation and flags it as an error.

However, DifuzzRTL and ProcessorFuzz still suffer from low-quality test stimuli due to their mutation schemes and corpus scheduling approaches. ProcessorFuzz and MorFuzz incur substantial time overhead because they rely on software simulators. Cascade avoids the overhead of output comparison by modifying its bug evaluation mechanism, but risks missing bugs as a trade-off. Moreover, all these frameworks are implemented at the software level and do not prioritize the speed of test case generation, which can be inefficient especially when verifying large and complex hardware designs such as CPU cores. In comparison, as will be detailed in the following sections, TurboFuzz redesigns the corpus scheduling technique, introduces a static-dynamic hybrid fuzzing scheme, and implements the entire fuzzing system in hardware, including test generation, online checking and coverage collection, thus significantly improving the overall fuzzing performance.

\subsection{Automatic test stimulus generation}
Previous work has focused on automatic test pattern generation (ATPG) in the domain of structural fault testing. The D-Algorithm \cite{D_algorithm} was one of the earliest systematic approaches that introduced D-notation to symbolically model the logical discrepancy between fault-free and faulty circuits, enabling the algorithmic derivation of input combinations that activate and propagate faults to observable outputs. 

As VLSI complexity grew, Built-In Self-Test (BIST) emerged \cite{BIST_Mc}, integrating test pattern generators and response analyzers directly on-chip to simplify post-manufacturing and in-field testing. The Linear Feedback Shift Register (LFSR) became a standard on-chip pseudo-random pattern generator capable of producing long-period sequences. Extensions such as pseudo-exhaustive testing achieved exhaustive coverage of local subcircuits. However, purely pseudo-random patterns proved ineffective for sequential circuits with state-dependent faults. To address this limitation, researchers introduced weighted random testing \cite{weighted_BIST}, Hybrid BIST \cite{Hybrid_BIST}, Scan Chain \cite{Scan_Chain}, and Design-for-Testability (DFT) \cite{DFT} techniques. These advances formed an integrated hybrid strategy combining random and deterministic test generation for higher coverage and efficiency. However, these works focused on post-silicon testing instead of pre-silicon functional verification, which is the main focus of TurboFuzz.

The concept of random test stimulation later influenced software testing and security research, leading to the emergence of fuzz testing. Bart Miller et al. first applied random inputs to assess the robustness of UNIX utilities \cite{Fuzz_Original}, inspiring decades of research on automated input generation. Subsequent advancements such as AFL and Microsoft SAGE introduced coverage-guided and constraint-based exploration, evolving fuzzing from pure randomness into feedback-directed test generation.

\subsection{FPGA Accelerated Verification}
FPGA-accelerated techniques for verification and debugging have been actively researched\cite{dessert,signalcat_2022,attia2020statemover,encore_fpga23} for years. Beyond conventional vendor-provided debugging tools, researchers have focused on leveraging specific FPGA features to speed up debugging. For example, StateMover \cite{attia2020statemover,sm_fpt,sm_trets} uses the FPGA readback technique \cite{readback} to create checkpoints and transfer design states to external software simulators for debugging. StateMover's reliance on human judgment for capturing hardware snapshots limits its effectiveness with complex designs, as choosing the right moment to take snapshots is crucial for narrowing down potential bug locations. A recent study \cite{encore_fpga23} addresses this limitation by automatically triggering snapshots when mismatches occur between the design under test and its reference model. However, previous work still relies on software-generated test vectors to drive the DUT, which often creates a performance bottleneck. TurboFuzz addresses this issue by developing a synthesizable fuzzer IP and coverage collection scheme that combines the dynamic self-checking capability offered by previous work, creating an end-to-end verification platform with FPGA acceleration.

\section{The TurboFuzz Framework Overview}

TurboFuzz is a hardware and software framework for fully-automated processor verification, the overall architecture diagram is shown in Fig.~\ref{Fig:overview}. The key characteristics and features of TurboFuzz include:

\hl{Hardware fuzzer}. The core of the TurboFuzz framework is the TurboFuzzer, a highly configurable hardware module that generates fuzzing test cases in two modes: direct mode, which produces tests with constrained-random stimulus; and mutation mode, which automatically varies stimulus based on coverage feedback to explore under-covered verification areas. Importantly, TurboFuzzer is fully synthesizable and implemented on an FPGA with minimal area overhead, enabling high-performance fuzzing.

TurboFuzzer also features a novel scheme named \textit{deepExplore} for optimal coverage convergence. This hybrid approach first uses direct testing with program samples that represent key features of the original benchmark, modified based on the SimPoint method\cite{Simpoint}. This achieves more extensive coverage of static and deterministic points while building high-quality seeds before transitioning to high-performance fuzzing to explore harder-to-reach states.

\hl{Automatic coverage instrumentation and collection}. The TurboFuzz framework automatically instruments synthesizable coverage points based on the Register Coverage metric — a widely used approach in recent hardware fuzzing works\cite{DifuzzRTL}, but with optimizations to eliminate unreachable coverage points, which improves effectiveness by better representing the unlimited design space with limited coverage points. This enhancement increases fuzzing efficiency for both TurboFuzzer and existing solutions. 

\hl{Fine-grained self-checking with FPGA acceleration}. The fuzzer-generated inputs feed into a differential checking module based on the \textsc{Encore} structure\cite{encore_fpga23}, which performs self-checking between the DUT and its reference model running concurrently on the same FPGA SoC. 
TurboFuzz also leverages the hardware snapshot capability that captures the complete state of the FPGA, including logic blocks, flip-flops, on-chip memories, and external devices such as DDR. Users can transfer these snapshots from the FPGA to analyze and debug issues using external software simulators like ModelSim on a host PC\cite{attia2020statemover,sm_fpt}.

\section{Hardware Architecture of the TurboFuzzer}\label{sec:fuzzer_arch}
\begin{figure}[tb]
  \centering
  \includegraphics[width=.4\textwidth]{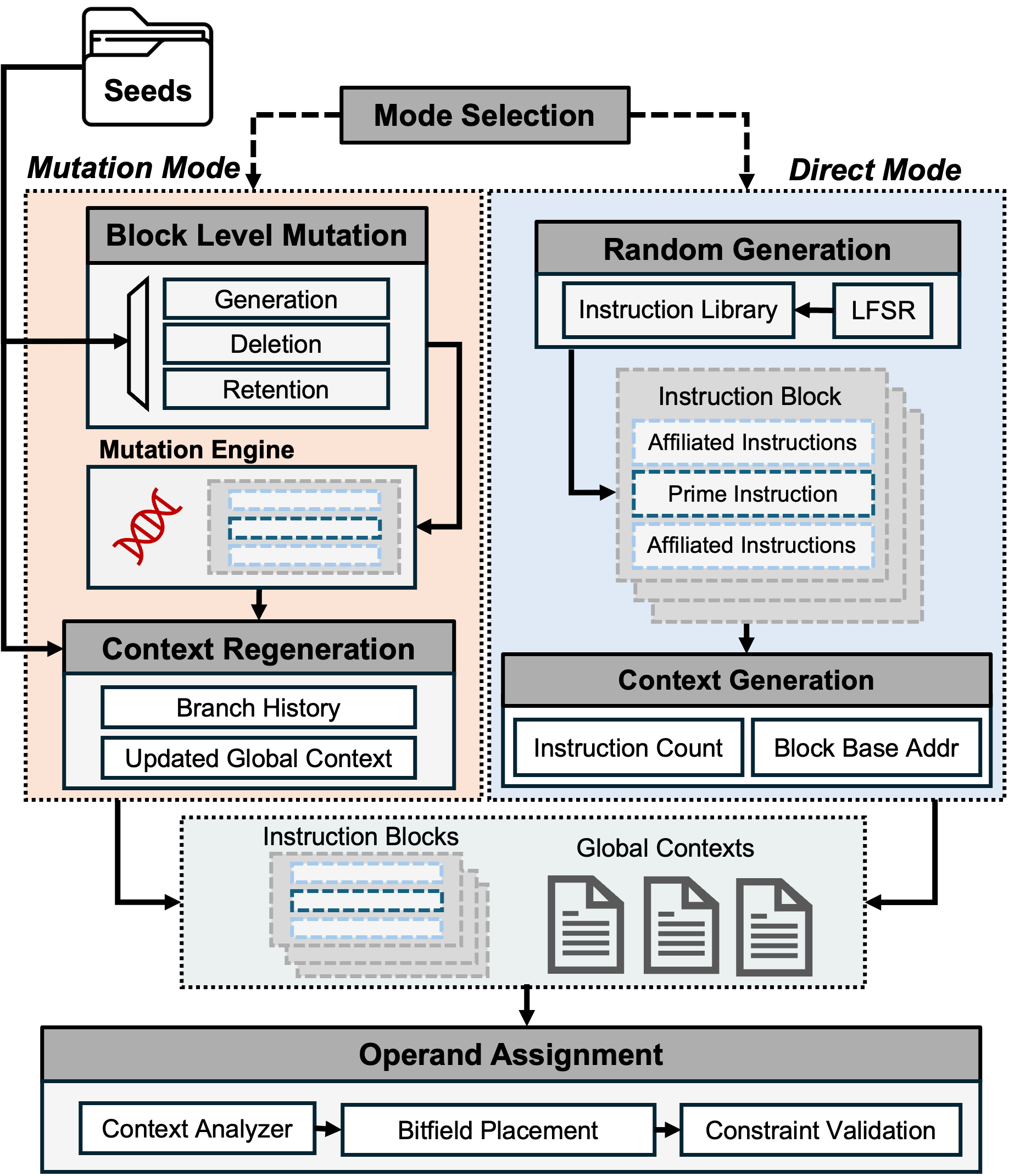}
  \caption{General architecture and workflow of TurboFuzzer.}
  \label{Fig:fuzzer_arch}
\end{figure}

This section introduces key concepts used in the fuzzing system design, followed by a detailed explanation of the TurboFuzzer design as shown in Fig.~\ref{Fig:fuzzer_arch}, including processor instruction generation, mutation, and optimization methods.

\subsection{Notions}

\subsubsection{Instruction block}
The instruction block is the basic unit for \textit{generating} instructions, consisting of a mandatory prime instruction and optional affiliated instructions. This is because some instructions require integration with others to meet architectural and testing requirements.

Affiliated instructions can be positioned before the prime instruction to establish prerequisites, or after to specify follow-up operations. In this paper, we use ``instruction block'' to refer to both standalone prime instructions and prime instructions with their affiliated instructions.

\subsubsection{Fuzzing iteration}
To verify modern CPUs effectively, multiple instructions must be run together to test specific states. The fuzzer's basic \textit{output} unit - an iteration - consists of multiple instruction blocks containing anywhere from tens to thousands of instructions.

\subsubsection{Seed and Corpus}
Seeds store valuable instruction sequences with metadata for mutation operations. Each seed's stimulus entry contains the instruction, its iteration position, control-flow status, and branch target position. This structure enables precise pattern reproduction while maintaining architectural context.

Seeds are stored in either FPGA on-chip BRAM (faster, limited capacity) or off-chip DDR memory (larger storage). The system randomly selects one seed per fuzzing iteration.

\subsection{Instruction generation}

\subsubsection{Seed selection}
During system initialization, the seed is populated with baseline instruction sequences. Throughout the fuzzing operations, the corpus manager maintains the corpus through two primary mechanisms: seed assembling with metadata packaging, and prioritizing coverage through state transition analysis. This prioritization quantifies each seed's coverage improvement contribution through architectural state transition analysis, subsequently organizing the seeds into a coverage-annotated graph structure that preserves execution path relationships while enabling efficient mutation path selection.

After seed selection, the fuzzer sends instructions from the selected seed to the mutation engine sequentially. For each state transition, the mutation engine chooses between direct mode or mutation mode based on a configurable probability, of which the default value for mutation mode is 7/16, and the direct mode is 9/16.

\subsubsection{Direct mode}
In direct mode, the fuzzer generates executable instruction sequences randomly without using coverage-guided mutation. Specifically, it activates the random generation module which employs a Linear Feedback Shift Register (LFSR) to generate controlled pseudorandom values for selecting prime instructions randomly from the instruction library, and setting the initial instruction count for each instruction block as a general guidance.

Within this guidance, the final length of each instruction block adaptively adjusts based on the prime instruction's architectural constraints and affiliated instruction requirements. This dynamic sizing ensures execution integrity by bundling instructions in a context-aware manner while preserving the random nature of generation.

The instruction library is a dynamically configurable repository that contains the complete RISC-V instruction set. Individual instruction subsets (such as RISC-V I, M, F, A, ZiCsr) are organized into categories, allowing them to be activated or deactivated through VIO configuration interfaces. The library can accommodate future ISA extensions or architectural changes through direct modifications or complete replacement.

Selected instructions contribute to building the architectural context of the current iteration. The global execution context metadata maintains two key parameters: the cumulative instruction count, which is the total instructions generated within the iteration, and the block base address, which is the memory-aligned starting address of each instruction block. When the iteration generation is completed, the global context metadata records the final instruction count and code segment boundary.

In addition to specific instruction types, the fuzzer also generates immediate operands and address values, including memory access addresses for load/store instructions, and relative branch addresses for control flow operations. For memory address generation, a user-configurable probability parameter with a default value of 3/4 determines whether to select data segment addresses or instruction segment addresses. The generation of relative branch addresses follows architectural constraints defined by the global context metadata.

To ensure valid branch execution, branch targets must align with instruction block boundaries. The fuzzer achieves this by selecting target addresses from pre-recorded block base addresses and calculating branch offsets using program counter values and pipeline latency adjustments.

The operand assignment module uses operand data to populate instruction-specific bit fields for creating executable instructions. Since different instruction types require unique operand mapping strategies, this module first analyzes the global context metadata to identify the current instruction type. After classification, it retrieves pre-generated immediate values and address operands from the fuzzing context module and maps these values into their appropriate bit fields according to the RISC-V specification.

Completed instructions are validated against architectural constraints before being queued for commitment to the instruction segment in the DDR. This ensures ISA compliance while maintaining the fuzzer's randomization properties through instruction format checking.

\subsubsection{Mutation mode} 
The key difference between mutation mode and direct mode is that it uses coverage-driven strategies to modify existing instruction sequences. It leverages the real-time coverage information from the DUT to guide targeted modifications, generating instructions that effectively explore uncovered states. 

The mutation mode comprises four main phases: 
\begin{enumerate}
    \item Coverage-aware seed selection from the corpus.
    \item Targeted instruction mutation using bit-flipping and operand substitution.
    \item Context regeneration with coverage-based constraints.
    \item Coverage-sensitive operand rebinding.
\end{enumerate}

Each phase dynamically adapts its transformation rules based on ongoing coverage patterns. This guides the fuzzing process toward unexplored execution paths while ensuring instructions remain architecturally valid.

The initial phase of mutation involves selecting optimal seeds from the corpus. The selection process uses a dual-strategy probabilistic mechanism with a configurable probability (default value 3/4), which guides the system to either prioritize seeds with high coverage improvement or perform random selection across the corpus. This random selection ensures all archived execution patterns are utilized, prevents high-potential seeds from being starved of resources, and avoids premature elimination of seeds before their coverage potential is realized. This balanced approach maintains an effective exploration-exploitation trade-off while preserving the evolutionary characteristics of the corpus through feedback.

Selected seeds move sequentially from corpus storage to the mutation engine, which performs three types of operations on instruction blocks: generation, deletion, or retention. These operations follow user-definable probabilities with default values of $3/16$, $11/16$, and $2/16$, respectively. The engine makes atomic modifications to complete instruction blocks while maintaining architectural consistency through hardware-enforced validity checks during the mutation process.

\textbf{The generation operation} is similar to direct mode, inserting new instruction blocks at the current execution point using stochastically generated parameters. This process updates the global context metadata to maintain synchronization between architectural states and instruction block boundaries.

\textbf{The deletion operation} removes instruction blocks by setting elimination flags in the control header of the iteration. This process preserves instruction stream integrity through pointer redirection while removing the block from subsequent processing stages.

\textbf{The retention operation} preserves the instruction block in the current iteration while logging critical global information to the execution context. This includes the instruction type, block generation status, and for control-flow instructions, the target block index from the current seed. For retained control-flow instructions, the system verifies that jump targets align with valid block starting addresses by checking the global address table. The recorded global information contributes to the context construction of the iteration.

Since retained instruction blocks already contain the prime-affiliated structure, they can skip the instruction injection phase used in direct mode. However, retention operations must still reconstruct context to maintain control-flow dependencies. For retained control-flow blocks, the system preserves their original jump targets when the referenced blocks remain in the current iteration. If target blocks are deleted, the system generates new relative offsets by randomly selecting from the global block address table, ensuring valid transitions to the remaining instruction block headers.

After instruction selection and context establishment, the unified operand assignment module fills instruction fields for both generated and mutated blocks. This shared component uses consistent bit-field manipulation rules across generation modes, placing precomputed immediate values and validated addresses into their designated positions as specified by the RISC-V ISA. Notice that for non-RISC-V ISAs, the generation and operand assignment modules within TurboFuzzer must be modified to fit the new ISA, while the mutation engine can remain unchanged.

\subsection{Optimizations on Control-Flow Instruction Generation}
Previous research has revealed that fuzzers face efficiency challenges despite generating many instructions. Not all generated instructions are actually executed, and of those that are executed, only a small portion are fuzzing instructions. Research shows that on average, only 19.3\% of generated instructions complete execution, while just 5.8\% of executed instructions are actual fuzzing instructions—the rest being non-randomized code like setup routines and exception handlers\cite{Cascade}. This significantly limits fuzzing efficiency since most generated fuzzing instructions go to waste.

\begin{figure}[tb]
  \centering
  \includegraphics[width=.48\textwidth]{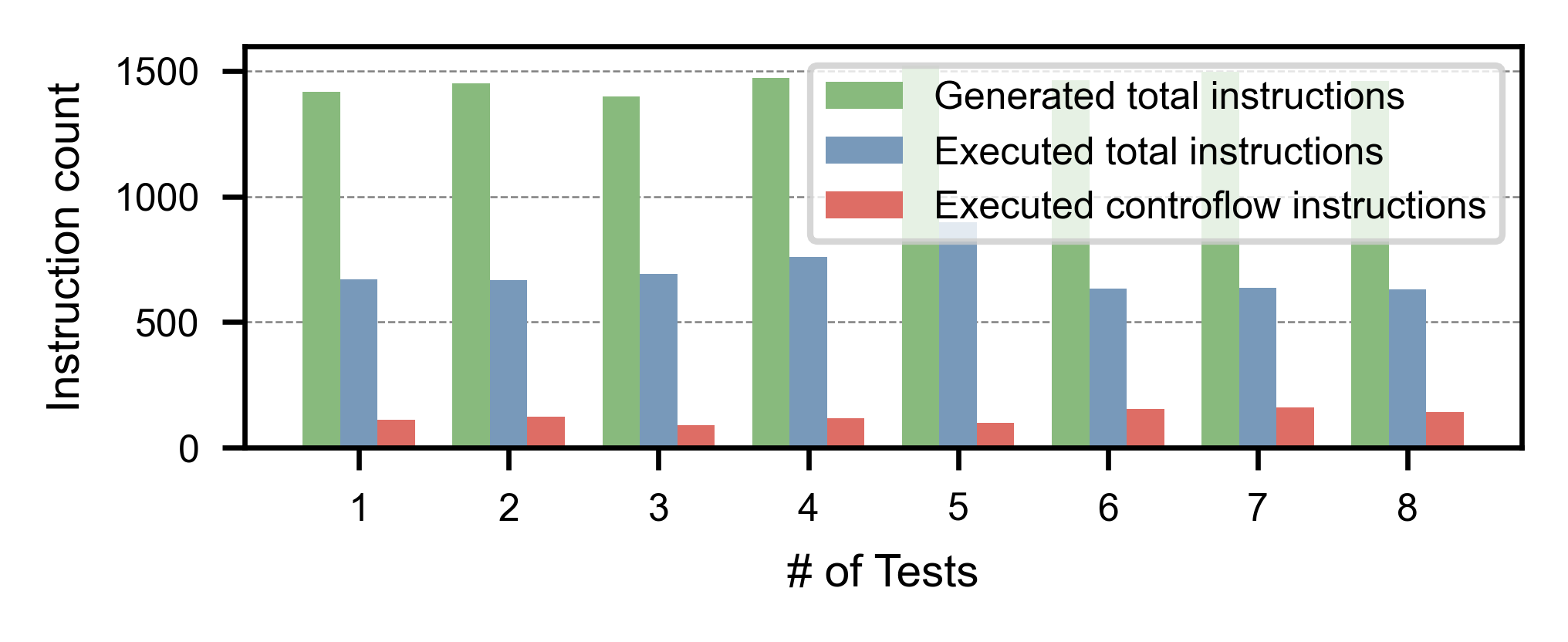}
  \caption{Proportion of executable instructions.}
  \label{Fig:control_flow_dist}
\end{figure}

To investigate the root cause of the issue, we analyzed the instruction types in programs generated by DifuzzRTL. As shown in Fig.~\ref{Fig:control_flow_dist}, green bars represent overall instructions, blue bars show executed instructions, and red bars indicate control-flow instructions within the executed set. The analysis reveals that control-flow instructions—including jumps (jar and jalr) and branches (bne and bge)—comprise more than 1/6 of all instructions. To study this pattern, we examined a constrained scenario that only allows backward jumps to prevent infinite loops from forward jumps with a 1:5 ratio of control-flow to non-control-flow instructions, matching the previous experimental observations. When jump targets have no spatial constraints, let $p$ denote the current instruction position, and $L$ denote the terminal position, then the jump space is between $[1, L-p]$. Assuming uniform distribution across the instruction space, the expected jump distance is as shown in \eqref{Eq:jump_distance}. 

\begin{equation}\label{Eq:jump_distance}
\text{E}_{\text{j}} = \frac{1 + (L - p)}{2}
\end{equation}

We observe that for iterations of large length and with early positioned jump instructions, i.e. $p << L$, the expected jump distance $E_j$ significantly exceeds the 5-instruction window between consecutive control-flow operations. This leads to extensive instruction skipping, and jumps often land directly in the iteration's final section and thus bypassing many intermediate instructions entirely.

It is worth noting that the effectiveness of fuzzing depends heavily on how many instructions are executed in each iteration. To improve fuzzing quality, we must maximize the number of executed instructions per iteration. Our methodology accomplishes this through three key mechanisms:

\textbf{Increased instruction counts}: We scale the instruction count per iteration to 4000 instructions, in contrast to traditional methods that generate only 20 to circa 1000 instructions. While longer sequences naturally increase the expected jump distance as analyzed earlier, our approach uses compensatory constraints to minimize efficiency loss.

\textbf{Control-flow jump range limitation}: By restricting jump targets to a bounded range, we increase the proportion of executed instructions per iteration.  

\textbf{Templates with execution guarantee}: When encountering an exception, the system attempts to handle it using predefined templates and resumes normal execution. If the exception cannot be resolved, the system moves on to the next iteration.

Generating larger instruction volumes requires extended generation and mutation processes. As described in Section~\ref{sec:fuzzer_arch}, comprehensive global context recording is essential for controlling jump addresses during instruction generation. However, extended recording can cause the global context buffer to grow linearly. While this is not problematic for software-based fuzzers, FPGA implementations must carefully balance LUT and BRAM utilization to maximize synthesis efficiency. Sequential processing of large instruction sets is also inefficient. To improve performance with a limited area budget, we implement pipelined processing within and between functional units, leveraging parallelism of the reconfigurable hardware to optimize throughput while maintaining architectural consistency.

To constrain control-flow jump ranges, we classify jump operations as either generated or retained, leveraging the global context buffer. Both types of instruction blocks record their global information, including their position within the iteration. Retained instruction blocks also preserve the index of their original jump target block.

For generated control-flow blocks, we limit the delta values within the fuzzing context module to keep jump targets within a small range. Preserved control-flow instruction blocks, however, remain unrestricted since their seeds come from previously generated stimuli. Because the jumps in generated instructions are already constrained, preserved instructions naturally have a low probability of making long-distance jumps. We deliberately maintain small probability for the overall randomness and comprehensiveness of the fuzzing system.

To establish optimized templates, we implement enhanced exception-handling mechanisms. For example, floating-point instructions require specific bit fields in the Floating-Point Control and Status Register (FCSR) to be enabled. When an instruction triggers an exception due to disabled bit-fields, our exception handler re-enables the relevant FCSR bit-fields and continues fuzzing rather than terminating the iteration. This approach ensures proper logging of the exception scenario, while preserving valuable debugging information. The fuzzer then resumes execution to maximize iteration efficiency and process subsequent instructions effectively.

Similar to the software fuzzer, we implement memory access restrictions through the fuzzing context module. This component manages symbol values to enforce our memory address requirements. For memory access operations, we distinguish between read and write types. Read operations have a configurable probability distribution: 1/4 chance of accessing the instruction region and 3/4 chance of accessing the data region. These probabilities can be adjusted by users. The data region contains random sequences generated by an LFSR, with a unique seed for each fuzzing iteration. To prevent issues from self-modifying code, write operations are confined to the data region only.

\subsection{Corpus Scheduling Optimization}
Corpus serves as a fuzzer's knowledge base and foundation for generating test cases. A well-designed initial corpus containing valid, diverse inputs, including both legitimate and edge-case data that cover different code paths, helps the fuzzer produce meaningful new inputs more quickly while reducing ineffective attempts. Corpus-based mutation strategies like AFL's bit-flip and block-replacement operations prove more efficient than purely random generation\cite{afl-fuzz,honggfuzz,libfuzzer-geralis}. Effective \textit{corpus scheduling}—the management of corpus entries—is therefore crucial for optimizing test-case generation and mutation. This scheduling involves two key aspects: prioritizing which seeds to use and efficiently updating the corpus through strategic addition and removal of entries.

Traditional fuzzing methods use simple first-in-first-out scheduling, replacing the oldest seed when the corpus reaches capacity. While this approach is straightforward to implement, it has a significant drawback: it may discard valuable older seeds that continue to increase coverage. Meanwhile, newer seeds might not improve coverage as effectively as existing ones. Therefore, seed selection should consider not just age, but also the ongoing effectiveness at discovering new paths of each seed.

\begin{figure}[tb]
  \centering
  \includegraphics[width=.48\textwidth]{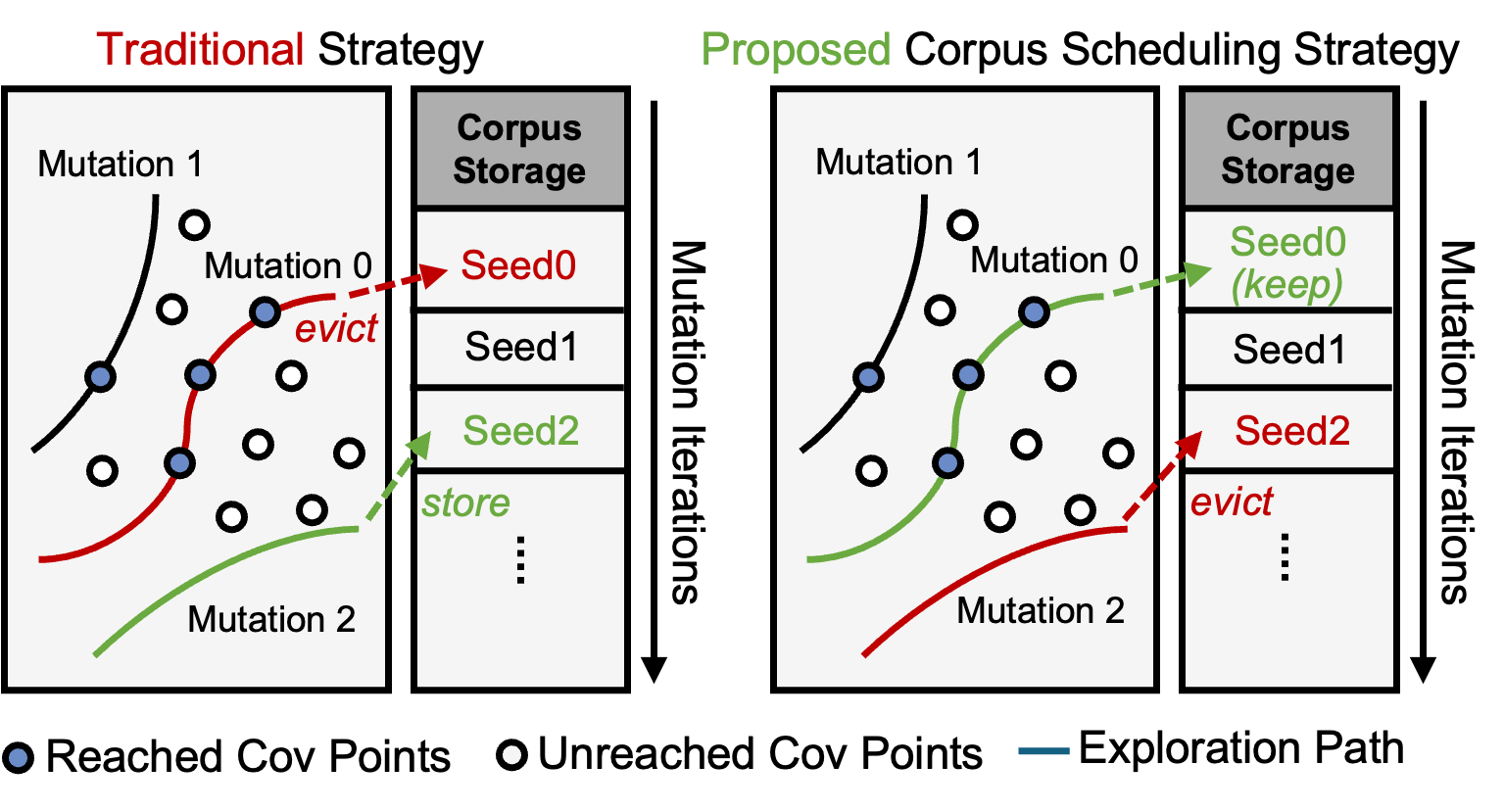}
  \caption{Comparison of the corpus scheduling strategies.}
  \label{Fig:corpus_schedule}
\end{figure}

Therefore, rather than removing seeds based on their age in the corpus, we prioritize keeping seeds that show promise in generating valuable test cases, as shown in Fig.~\ref{Fig:corpus_schedule}. To achieve this, we designed a corpus scheduling strategy based on coverage analysis and implemented it in the TurboFuzzer. Along with the original seed content, we track the coverage increment of each seed in the corpus, which informs our scheduling strategy:

\begin{itemize}
    \item \textbf{In generation mode}: New test cases are added to the corpus only if they improve coverage and there is available storage capacity. When the corpus reaches full capacity, we compare the new test case against existing seeds and replace the seed with the lowest recorded coverage improvement.
    \item \textbf{In mutation mode}: When we mutate an existing seed from the corpus and run it on the DUT, we measure its coverage and update the coverage improvement of that seed in the corpus.
\end{itemize}

\section{Exploring Broader Design Spaces}
The goal of a typical fuzzing system is to explore corner cases that may trigger security issues or vulnerabilities. However, from a coverage convergence perspective, this represents only a subset of the overall coverage space, while the remainder consists of deterministic design behaviors with natural data or control dependencies. These dependencies represent execution scenarios that are straightforward for ordinary software programs such as processor benchmarks. Due to the inherent tendency of fuzzers to generate random instruction streams, producing deterministic programs becomes relatively difficult, even with coverage-guided mutation. As a result, coverage points that are easily triggered by normal, non-random programs can be inefficiently reached by fuzzers.

To further explore a broader range of design spaces, we developed a hybrid approach called \textit{deepExplore} that combines direct tests and fuzzing to leverage the strengths of both approaches. The direct test cases are derived from standard CPU benchmarks such as coremark\cite{gal2012exploring}, dhrystone\cite{weicker1984dhrystone}, and microbench\cite{yy2021microbench}. However, running these benchmarks—even with FPGA acceleration—can take hours to complete since each benchmark contains hundreds of millions of instructions.

Additionally, since most of the benchmarks are constructed for performance evaluation rather than functional verification or coverage convergence, they normally contain large volumes of repetitive instructions such as tens of thousands of iterations within a single loop. In contrast, our primary interest lies in exploring diverse execution paths. Hence, it suffices to extract representative instruction fragments from benchmarks for inclusion in the corpus to accelerate coverage convergence. 

Software programs often exhibit recurring behaviors during execution. The SimPoint technique\cite{Simpoint} leverages this feature by identifying repetitive patterns to expedite the simulation of program execution. Specifically, SimPoint divides the program's execution into intervals and represents each interval as a Basic Block Vector (BBV), which captures the frequency of execution for each basic block within the interval, providing an architecture-independent profile of the program's behavior. By applying clustering algorithms, such as k-means, to BBVs, SimPoint groups intervals with similar behavior into clusters. From each cluster, a representative interval, termed a \textit{simulation point}, is selected. By simulating only representative points from each cluster and applying appropriate weights, SimPoint can efficiently estimate overall program performance.

Therefore, we develop two stages in \textit{deepExplore}, the first stage extracts representable program intervals from benchmarks, and runs them in hardware to build high-quality seeds, which guide the fuzzer toward generating similar structural programs. Then it switches to the second stage, which employs the fuzzer for more extensive testing. 

Specifically, we first use the traditional SimPoint tool to extract multiple representative instruction intervals. These intervals achieve approximately 80\% overall coverage of the benchmark while consuming less than 1\% of its total execution time. However, these intervals cannot be directly executed on the DUT. Each interval requires its execution context, specifically the architectural state of the processor to construct the necessary initialization instructions, including the General Register File (GRF), Floating-Point Register File (FRF), and CSR. We use this context to construct initialization instructions that, combined with the instruction interval, create a complete executable instruction block.

To integrate the extracted instruction intervals with the fuzzer and enhance test generation, the hardware system continuously monitors interval execution. It identifies intervals that yield significant coverage improvements and marks them for future use. These marked intervals, which represent more structured behavior patterns, increase the likelihood of reaching similar deterministic states and achieving higher coverage.

After executing all instruction intervals, we preserve the register dependency structure of the marked intervals while mutating their initialization states, including memory addresses and immediate values. By executing these lightly mutated intervals, we can explore state spaces that are similar to, yet distinct from, the original intervals. Throughout this process, we monitor how each mutated interval improves coverage based on its new initialization data. The fuzzing system analyzes this feedback to refine the initialization parameters for subsequent execution iterations. This iterative process continues until coverage improvements begin to plateau.

\section{Coverage Instrumentation}\label{sec:cov_instrument}
Existing fuzzers for hardware designs base their coverage instrumentation methods on control registers. The coverage instrumentation algorithm first identifies all multiplexers within a design module. For each multiplexer, it then recursively traces backward through connected registers until reaching the module boundary. During this trace-back process, any registers encountered are designated as control registers for that multiplexer. After completing this procedure for all multiplexers, the algorithm produces a complete set of control registers for the module. These control registers store the coverage point indexes for the module. The process can be performed iteratively throughout the design hierarchy to achieve coverage instrumentation for the entire design.

However, combining all control registers from every module would require an exponential buffer size to represent the complete set of coverage points. To address this, previous work constrains the combined bit-width of control registers within a predefined maximum state size. When the total bit-width exceeds this threshold, each control register undergoes a randomized bit-shift operation within the allowed range, with zeros filling any empty positions. The system then performs sequential XOR operations on these intermediate signals to generate the new coverage point index.

However, the existing instrumentation method has two major flaws. The first is modulo bias. The number of coverage points is directly tied to the number of multiplexers in a module. This means modules like multipliers and dividers, which contain many frequently toggled multiplexers, would generate an unusually high number of coverage points. Additionally, these modules end up providing stronger feedback to the fuzzing system, causing it to generate more test cases targeting them. As a result, the fuzzer focuses too heavily on arithmetic units while potentially neglecting other critical components like control paths. 

Second, this method can create unreachable coverage points. The original algorithm's randomized shifting of control registers with zero-padding leads to XOR operations where multiple padded zeros produce a zero result. This creates coverage points that don't correspond to any actual control register, effectively introducing idle points into the system.

\begin{figure}[tb]
  \centering
  \includegraphics[width=.48\textwidth]{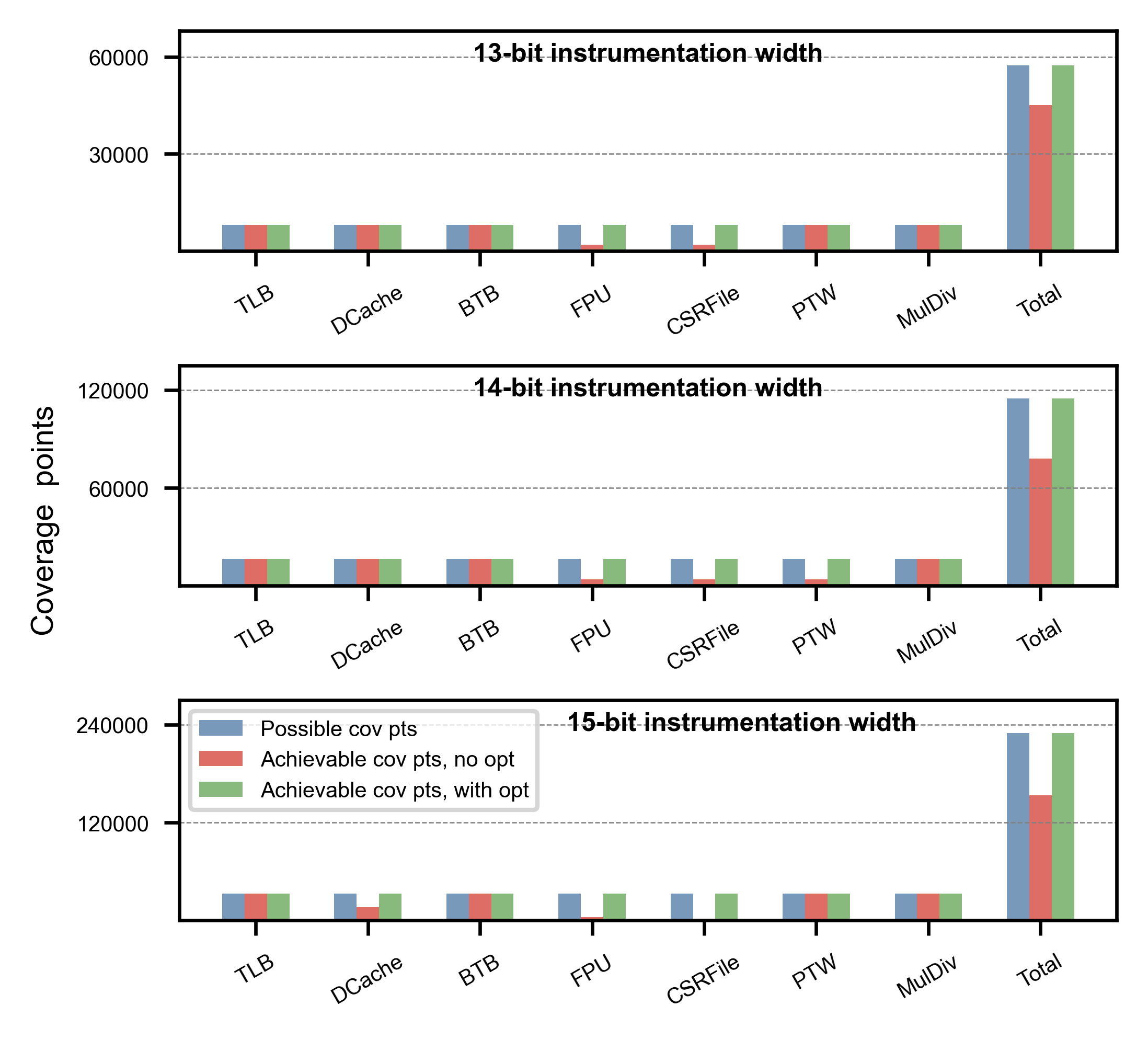}
  \caption{Distribution of the instrumented coverage points and the achievable ones with and without optimizations.}
  \label{Fig:cover_instrumentation}
\end{figure}

To demonstrate this issue, we applied the existing coverage instrumentation method to RocketChip\cite{rocket} and analyzed both the instrumented and unreachable coverage points across several major modules and the entire design, using various maximum state sizes, as shown in Fig.~\ref{Fig:cover_instrumentation}. The results show that only 76.8\%, 65.5\%, and 61.4\% of total coverage points are reachable across different configurations—with larger numbers of instrumented points correlating to lower achievability. Moreover, modules such as FPU, CSRFile, and PTW demonstrate particularly poor coverage point accessibility, with only a small fraction being achievable.

To improve coverage instrumentation efficiency, we developed two optimization techniques. First, to address the bias issue in the original algorithm, we added an auxiliary register alongside the original register that stores the triggered instrumentation coverage points $N_{cov}$ for each module. This auxiliary register performs shift operations on the original value to increase or decrease $N_{cov}$. Instead of using the original $N_{cov}$, the fuzzing system uses shifted value as the metric for coverage feedback, which provides straightforward yet effective control over each module's feedback intensity. For example, we can reduce the $N_{cov}$ value of the MulDiv module to decrease its weight relative to other modules.

In practice, we implement this strategy during the FIRRTL instrumentation stage by explicitly specifying left or right shifts for each module, which strengthens or weakens its coverage feedback. To further enhance flexibility, we allow users to specify individual top-level modules for coverage instrumentation, instead of requiring instrumentation across the entire design, enabling targeted coverage monitoring of specific design modules.

To address the issue of unreachable coverage points, we developed an alternative to the traditional method of randomized shifting and zero-padding. Our approach eliminates randomness by arranging control registers sequentially within the predefined threshold $maxStateSize$. We maintain a record of each register's positional offset during sequential arrangement. When the sum of the current offset and a control register's width exceeds the threshold, we roll back the new offset according to equation \eqref{Eq:instrument_offset}, where $W_{ctrl}$ represents the control register's bitwidth.
\begin{equation}\label{Eq:instrument_offset}
\text{new\_offset}=(\text{last\_offset} + \text{W}_{\text{ctrl}}) ~\%~\text{maxSateSize}
\end{equation}

The coverage instrumentation results are also shown in green bars in Fig.~\ref{Fig:cover_instrumentation}. As the proposed new method eliminates the potential empty states, all instrumented coverage points can be practically achieved, thus improving resource utilization especially for hardware implementation.

\section{Evaluation}
\subsection{Experimental Settings}
We implement TurboFuzz on the Fidus Sidewinder board, which contains an AMD Zynq UltraScale+ XCZU19EG FPGA and 32GB DDR4 memories. The FPGA image is generated using Vivado 2020.2. We first evaluate the performance improvements of the proposed optimizations, including optimized coverage instrumentation, enhanced control-flow instruction generation, corpus scheduling, and the \textit{deepExplore} method. Then we combine all the optimizations and evaluate the performance improvements and overhead of the proposed TurboFuzz hardware framework against two state-of-the-art software fuzzing systems DifuzzRTL and Cascade, both targeting processor verification. In these experiments, we use the 64-bit in-order RISC-V Rocket core as the DUT running at 100MHz. We then extend the evaluation to include another widely used single-issue RISC-V core CVA6\cite{cva6}, and the out-of-order superscalar RISC-V core BOOM\cite{boom} to demonstrate the performance improvements when catching real-world bugs.

\subsection{Enhanced Coverage Instrumentation}
As detailed in Section~\ref{sec:cov_instrument}, we propose a new coverage instrumentation method that eliminates the potential unreachable empty states, which is critical for efficient hardware implementation to save resource usage. This is a universal method and can also be beneficial for conventional software-based fuzzing approaches because it better utilizes limited coverage points to represent the unlimited design space, enabling broader functionality coverage without modifying the fuzzer itself.

\begin{figure}[tb]
  \centering
  \includegraphics[width=.48\textwidth]{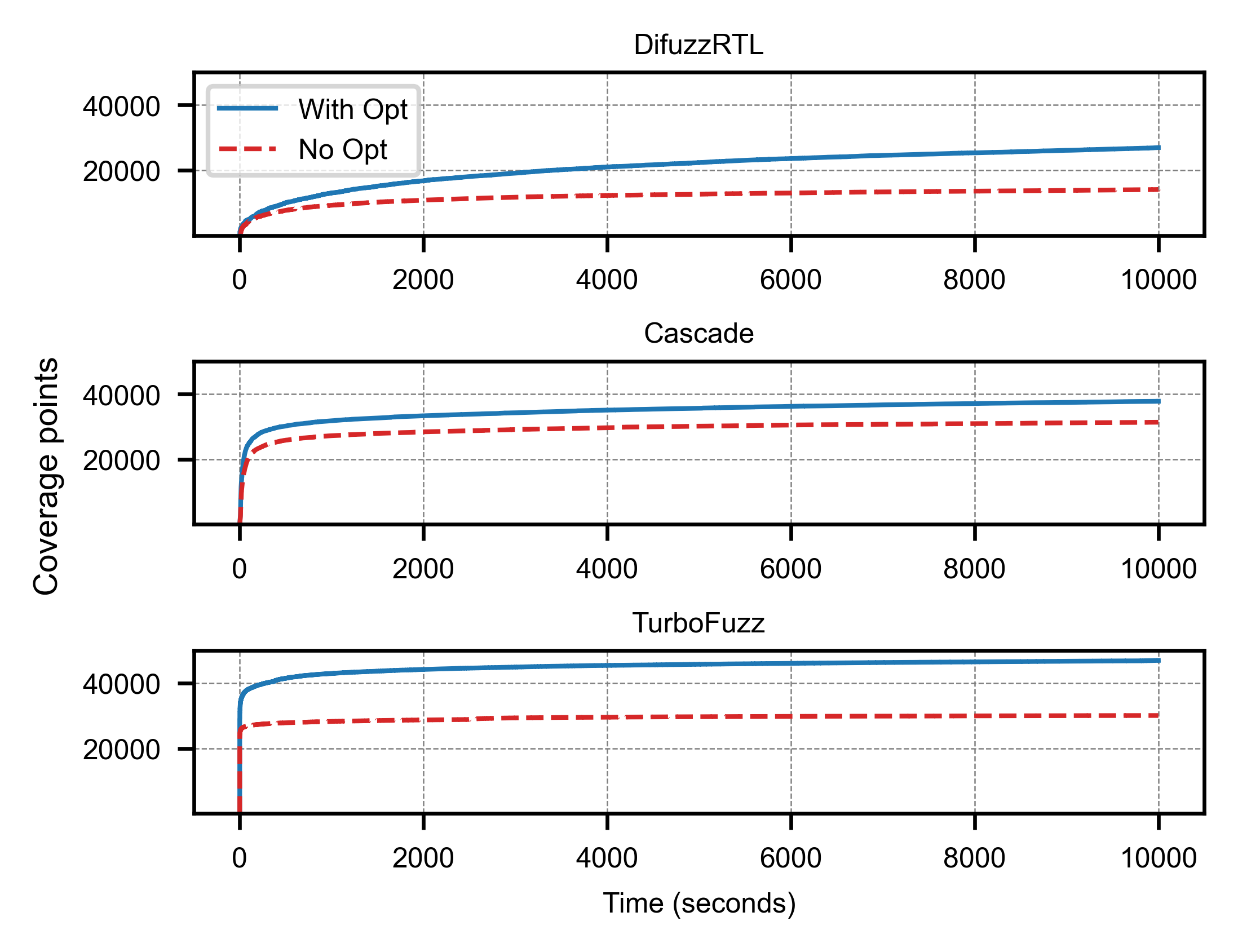}
  \caption{Comparison of the coverage improvements with proposed instrumentation method.}
  \label{Fig:cov_instrument_fuzz}
\end{figure}

As shown in Fig.~\ref{Fig:cov_instrument_fuzz}, we compare the reachable coverage points across all three fuzzing methods, using both their original instrumentation techniques and our proposed new techniques. The results demonstrate improvements in both the total number of reachable coverage points and coverage convergence speed. In our experiments, the maximum coverage points for DifuzzRTL, Cascade, and TurboFuzz increased by $1.91\times$, $1.21\times$, and $1.56\times$, respectively. Therefore in the following experiments, we will use the proposed coverage instrumentation methods for all fuzzing systems.

\subsection{Enhanced Instruction Generation}
\begin{figure}[tb]
  \centering
  \includegraphics[width=.48\textwidth]{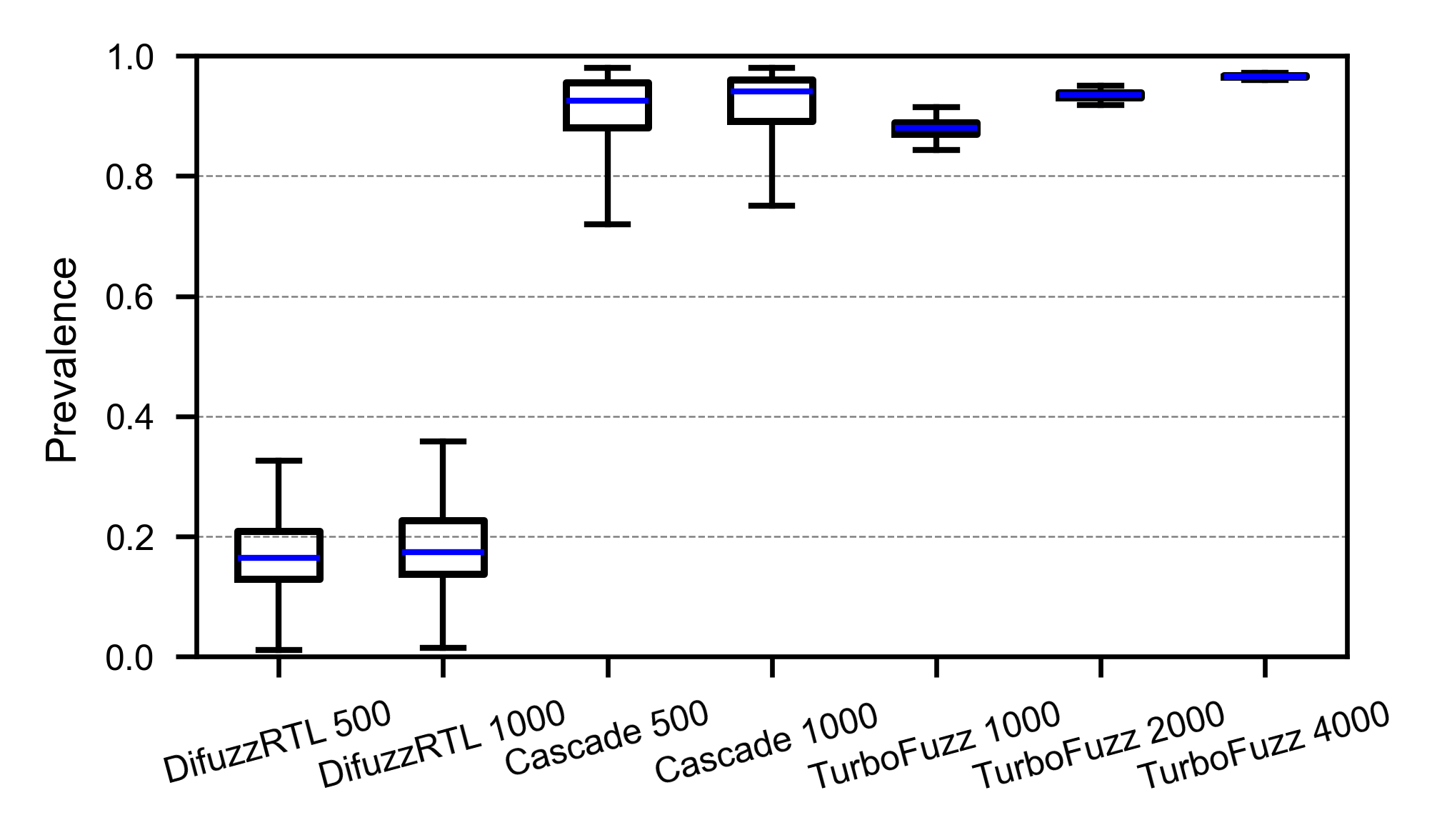}
  \caption{Prevalence comparison between different fuzzing methods with various configurations of instruction generation per iteration.}
  \label{Fig:prevalence}
\end{figure}

TurboFuzz optimizes instruction generation, particularly for control-flow instructions, to generate more high-quality instructions that can be executed. To quantify this improvement, we use the concept of \textit{prevalence} from previous work\cite{Cascade}, which is defined as the ratio of fuzzing instructions to total executed instructions, and compare prevalence values across different fuzzing schemes in Fig.~\ref{Fig:prevalence}.

For DifuzzRTL, less than 20\% of the executed instructions are fuzzing instructions, with the remainder being non-random and routine codes. In this case, even when configured to create more instructions, most of them are actually wasted. In comparison, Cascade is designed specifically for high prevalence, achieving an average prevalence of 0.93, with minimum and maximum values of 0.72 and 0.98, respectively. TurboFuzz, with its optimizations for control-flow instruction generation, achieves an even higher average prevalence of 0.97 while maintaining a consistent range among multiple tests, ranging between 0.96 and 0.97 with 4000 instructions per iteration.

\subsection{Enhanced Corpus Scheduling}
\begin{figure}[tb]
  \centering
  \includegraphics[width=.48\textwidth]{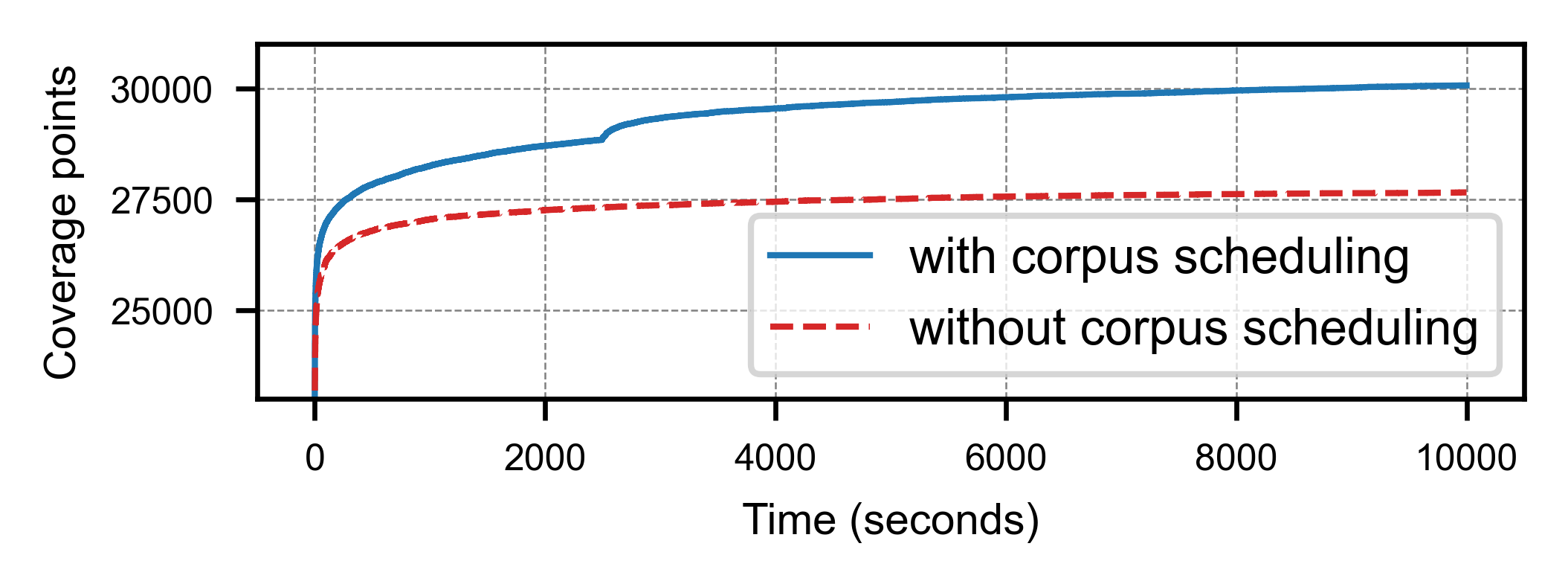}
  \caption{Coverage with different corpus scheduling techniques.}
  \label{Fig:corpus_sche}
\end{figure}

TurboFuzz implements an optimized corpus scheduling technique to maintain promising seeds for further mutation, rather than naively replacing seeds in a FIFO manner like software fuzzers. We compare the coverage status of the TurboFuzz framework with this feature enabled versus disabled, as shown in Fig.~\ref{Fig:corpus_sche}. To isolate the effect of corpus scheduling, we exclude software fuzzer comparisons to eliminate any performance benefits from hardware acceleration.

The results demonstrate significant coverage improvements when corpus scheduling is enabled. With a one-hour time budget, the optimized version achieves 7.5\% better coverage than the conventional approach. Similarly, to reach a target of 27500 coverage points, the optimized approach shows a $17.7\times$ speedup. 

Interestingly, we also observed a distinct coverage jump in the optimized implementation after generating circa 400 million instructions in less than one hour of execution. This jump indicates the fuzzer successfully reached an underexplored space, likely due to better seed selection in the corpus-a behavior not observed when corpus scheduling was disabled.

\subsection{Effectiveness Evaluation of deepExplore}
\begin{figure}[tb]
  \centering
  \includegraphics[width=.49\textwidth]{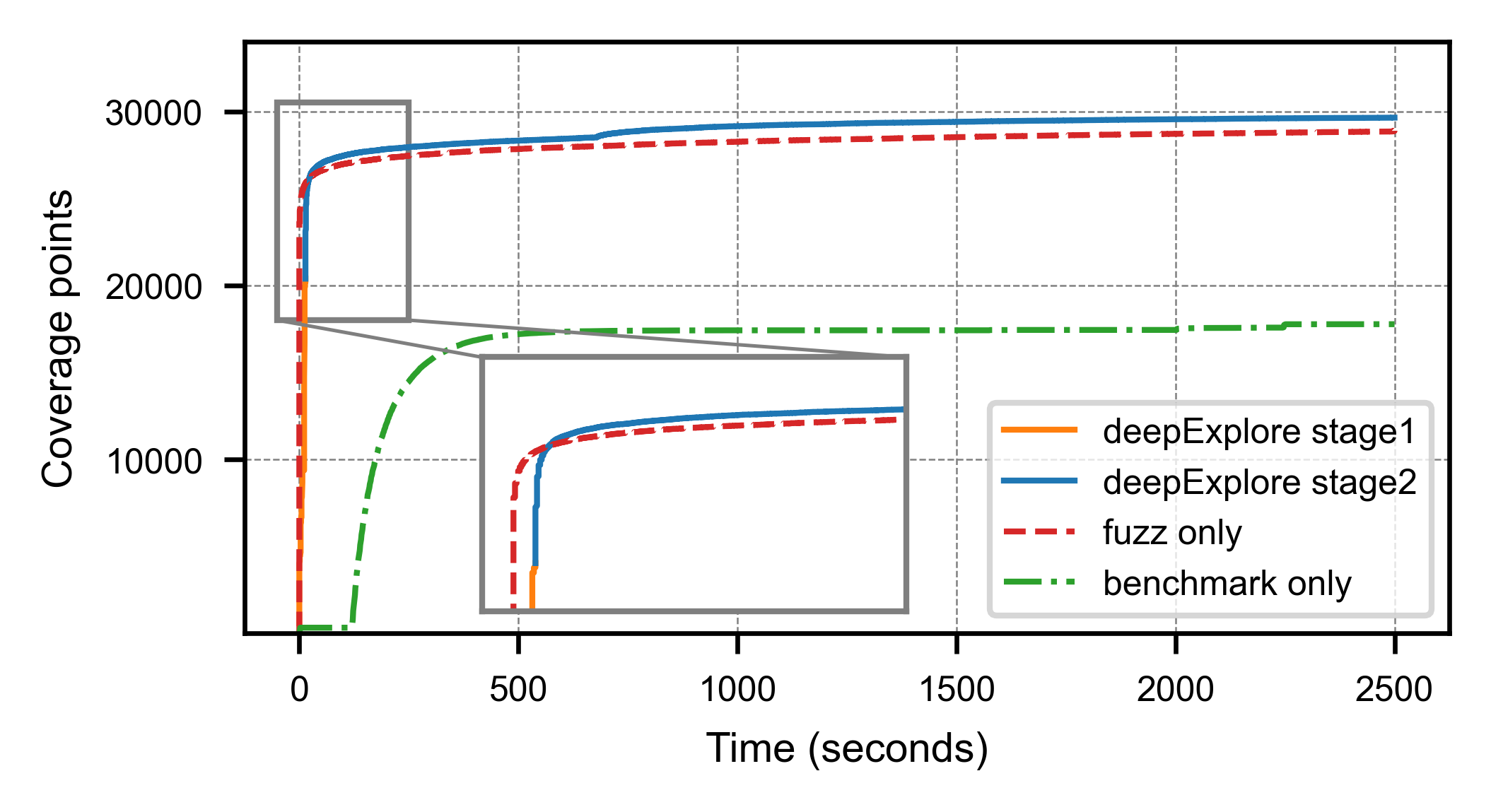}
  \caption{Coverage convergence with deepExplore. Stage1 denotes running sampled program intervals and building high-quality seeds, stage2 denotes using hardware fuzzing for high-performance tests.}
  \label{Fig:deepexplore}
\end{figure}

TurboFuzz embodies the newly proposed \textit{deepExplore} method, which is essentially a hybrid approach that can potentially explore a broader range of verification spaces. To demonstrate the effectiveness of \textit{deepExplore}, we measure the coverage increment of the TurboFuzz system with \textit{deepExplore} enabled, disabled (i.e. only perform fuzzing), and FPGA-based CPU benchmark execution without fuzzing. The results are shown in Fig.~\ref{Fig:deepexplore}.

In comparison to only running CPU benchmarks, TurboFuzz achieves coverage improvements by up to $1.67\times$, demonstrating that non-randomized programs cannot efficiently explore the verification space on their own. Additionally, \textit{deepExplore} approach covers 2.6\% more states than pure fuzzing. Though this improvement seems modest, it is crucial for achieving full coverage since convergence rates slow dramatically at higher coverage levels, particularly for hard-to-reach corner cases. This highlights a key limitation of pure fuzzing: while it excels at generating random test patterns, it struggles to create the structured patterns commonly found in standard programs.

It is also worth noting that the cross-point between the fuzz-only and deepExplore approaches occurs at approximately 22 seconds, as shown in the zoomed-out region of Fig.~\ref{Fig:deepexplore}. Before this point, running TurboFuzzer alone is faster since \textit{deepExplore} requires additional time to run extracted program intervals and construct extra seeds in the corpus. However, \textit{deepExplore} ultimately improves overall coverage by creating structured programs that pure fuzzing struggles to generate.

\subsection{Performance evaluation}
\begin{figure}[tb]
  \centering
  \includegraphics[width=.49\textwidth]{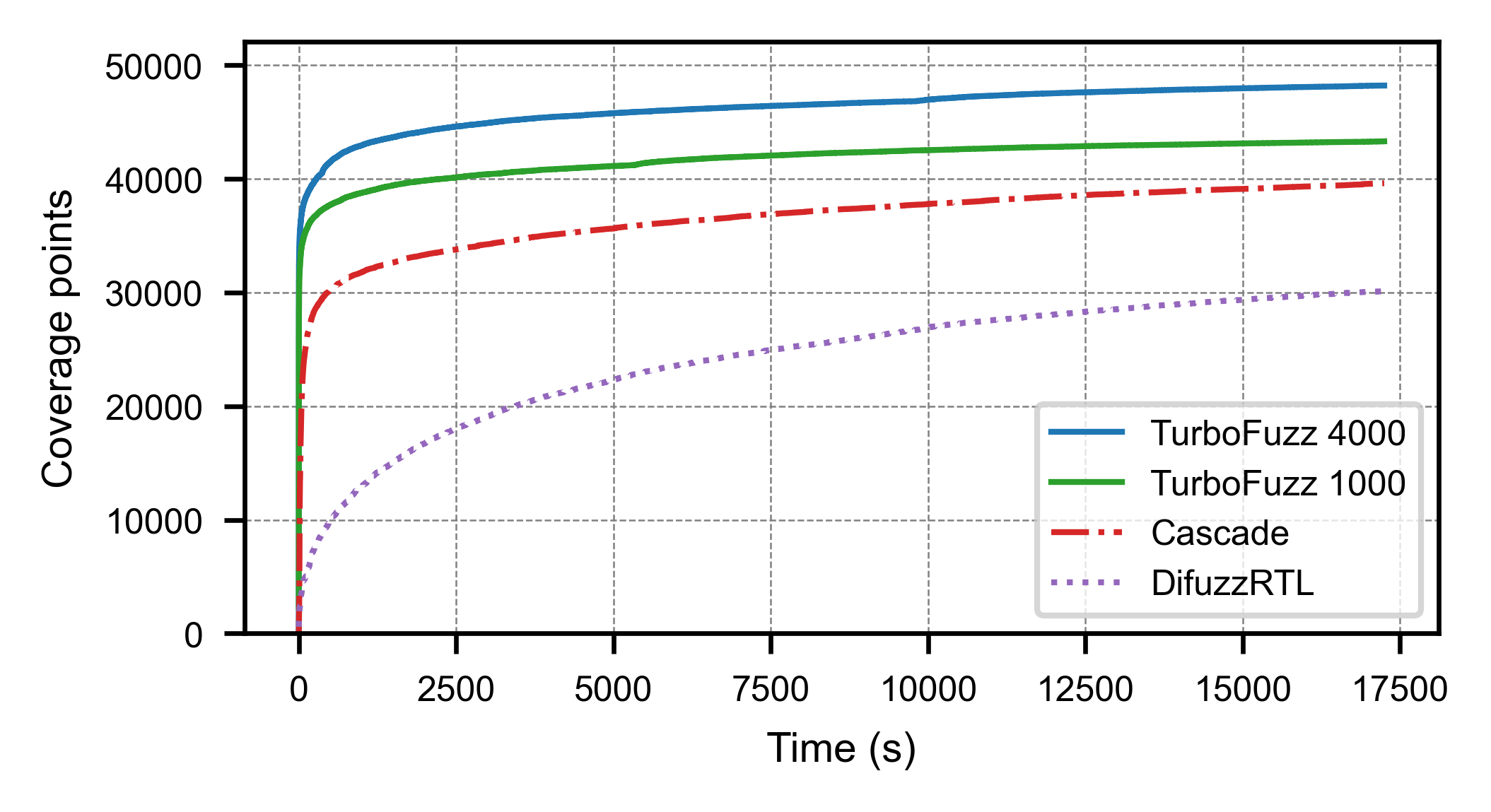}
  \caption{Comparison of coverage convergence performance between different fuzzing methods.}
  \label{Fig:perf_compare}
\end{figure}

Altogether, we combine all the aforementioned features and optimizations, and compare the performance for the three fuzzing methods, as shown in Fig.~\ref{Fig:perf_compare}. We maintain the default instruction count settings for the software-based fuzzers, and configure TurboFuzz to generate 1000 and 4000 instructions per iteration. 

For TurboFuzz, increasing instructions per iteration improves coverage collection performance by up to $1.11\times$ for the same running time, because of the high prevalence as shown in Fig.~\ref{Fig:prevalence}. Compared to software-based fuzzers, TurboFuzz achieves consistently better coverage across the entire time spectrum. When running for one, two and four hours, TurboFuzz outperforms Cascade in coverage collection by $1.26\times\sim1.31\times$, and DiffuzRTL by $1.64\times\sim2.23\times$. The improvements are even more dramatic for specific coverage targets. For instance, TurboFuzz reaches 35000 coverage points in just 14 seconds, indicating a $278\times$ speed-up compared to Cascade's 3,893 seconds. DifuzzRTL fails to achieve this coverage even after five hours of runtime.

Notice that DifuzzRTL can also offload DUT execution to FPGA. In this case, we measure the fuzzing performance of all three methods, as shown in Table~\ref{tab:fuzzing-performance}. Although FPGA accelerates DUT execution, data transmission between the host and FPGA creates a performance bottleneck, resulting in slower overall fuzzing speed. In contrast, TurboFuzz implements the fuzzer, DUT execution, and coverage collection entirely in hardware, achieving the highest fuzzing speed and instructions executed per time unit.

The performance improvements demonstrate the significant benefits of implementing the entire fuzzing system in hardware, enabling faster coverage convergence and allowing verification engineers to identify corner cases more efficiently.

\begin{table}[t]
\centering
\caption{Fuzzing Performance Comparison.}
\label{tab:fuzzing-performance}
\begin{tabular}{l@{\hskip 2pt}c@{\hskip 5pt}c}
\toprule
\textbf{Fuzzer} & \textbf{Fuzzing Speed (Hz)} & \textbf{Executed Inst per Second} \\
\midrule
\makecell[l]{DifuzzRTL\\(with FPGA)} & 4.13    & 728       \\
\midrule
Cascade & 12.80   & 2489     \\
\midrule
TurboFuzz      & 75.12   & 309676    \\
\bottomrule
\end{tabular}
\end{table}

\begin{table*}[t]
  \caption{Comparison on Bug Identification Performance.}
  \label{tab:bugs}
  \centering
  \footnotesize
  \setlength{\arrayrulewidth}{0.4pt}
  \setlength{\tabcolsep}{4pt}
  \renewcommand{\arraystretch}{1.15}

  \newcolumntype{L}[1]{>{\raggedright\arraybackslash}p{#1}}
  \newcolumntype{P}[1]{>{\raggedright\arraybackslash}p{#1}}

  \begin{tabular}{c|c|L{8.8cm}|P{2.2cm}|P{2.2cm}|P{2.2cm}}
    \toprule
    \textbf{Design} & \textbf{ID} & \textbf{Bug Description} &
    \textbf{SW Time (s)} & \textbf{HW Time (s)} & \textbf{Acc.\ Ratio} \\
    \midrule
    \multirow{10}{*}{\centering CVA6}
      & C1  & Incorrect setting of DZ flag for 0/0 division                                   & 39.53  & 1.03 & 38.54 \\ \cline{2-6}
      & C2  & Incorrect \texttt{fflags} set when \texttt{fdiv} divides by $\infty$            & 701.95 & 1.48 & 474.08 \\ \cline{2-6}
      & C3  & Wrong handling of Invalid NaN-boxed single-precision \texttt{fdiv}              & 931.30 & 1.63 & 571.69 \\ \cline{2-6}
      & C4  & Same as C2 (double-precision)                                                   & 445.28 & 1.31 & 339.39 \\ \cline{2-6}
      & C5  & Double-precision multiplication yields wrong sign when rounding down            & 35.64  & 1.03 & 34.68 \\ \cline{2-6}
      & C6  & Duplicate of C3 (another stimulus)                                             & 442.63 & 1.31 & 337.40 \\ \cline{2-6}
      & C7  & Co-simulation mismatch when reading \texttt{stval} CSR                         & 19.48  & 1.01 & 19.28 \\ \cline{2-6}
      & C8  & RV32A enabled without RV64A fails to raise exception                            & 581.21 & 1.42 & 408.87 \\ \cline{2-6}
      & C9  & \texttt{fdiv} returns $\infty$ when dividing by 0                               & 610.81 & 1.42 & 430.55 \\ \cline{2-6}
      & C10 & Division of $+0$ by a normal value results in $-0$                              & 844.18 & 1.58 & 535.64 \\
    \hline
    \multirow{2}{*}{\centering BOOM}
      & B1  & Floating-point rounding mode not working correctly                              & 457.99 & 1.31 & 348.71 \\ \cline{2-6}
      & B2  & FP instruction with invalid \texttt{frm} does not raise exception               & 358.60 & 1.24 & 289.36 \\
    \hline
    Rocket & R1 & Executing \texttt{ebreak} does not increment \texttt{minstret}              & 18.22  & 1.01 & 17.98 \\
    \bottomrule
  \end{tabular}
\end{table*}

TurboFuzz incorporates dynamic self-checking capabilities on the FPGA, enabling fine-grained checking of internal signals and key registers. Beyond the Rocket core, we evaluated TurboFuzz on two other widely adopted RISC-V cores: CVA6 and BOOM. For each DUT, we identified previously reported issues and compared how quickly software fuzzers versus TurboFuzz could trigger these issues, calculating the acceleration ratio as shown in Table~\ref{tab:bugs}. During testing, TurboFuzz also discovered two new issues, labeled C9 and C10 in this table.

TurboFuzz not only identifies all known issues and finds new ones, but also discovers them significantly faster, achieving geometric mean acceleration ratios of $194\times$ for CVA6 and $317.7\times$ for BOOM. It is worth noting that TurboFuzz demonstrates even greater acceleration advantages, up to $571\times$ when detecting hard-to-reach bugs that typically take hours to trigger. Through its FPGA-based hardware acceleration and fine-grained self-checking capabilities, TurboFuzz successfully catches those issues within two seconds.

\subsection{Overhead analysis}
We evaluate the area overhead of the TurboFuzz framework, compared against the vendor-provided logic analyzer tool ILA\cite{ila}, which requires users to specify signals to monitor during run-time. The resource usage data and its portion of the available FPGA hardware are summarized in Table~\ref{tab:fpga_utilization}. The DUT column shows the resource usage of the DUT together with instrumented cover points, while the TurboFuzz column shows the area of the entire framework, including the Fuzzer IP, and online differential checking system, but excludes the DUT and the instrumented cover points to ensure a fair comparison with ILAs. ILA uses on-chip memories for trace buffering, and we study a tracing depth of 1024 (config1) and 65536 (config2).

It can be seen that the TurboFuzzer IP consumes a moderate 13\% logic resources and 18\% on-chip memories, while the entire TurboFuzz framework takes 17\%, 23\% and 13\% of FPGA area in terms of LUT, BRAMs and Registers, respectively. 

In comparison, ILA uses $2.05\times$ and $2.55\times$ more area in BRAM than TurboFuzz, because the area usage of ILA is determined by tracing depth, especially for BRAMs which store debugging information. In addition, ILA requires complete design recompilation when adding/removing signals, whereas TurboFuzz's hardware snapshot captures the entire design status for debugging without this requirement.

We also evaluate the area usage and maximum frequency of TurboFuzz across different coverage instrumentation configurations. Typically, more instrumented cover points capture more state but require more area. In our experiments, we test three configurations: \texttt{cov1}, \texttt{cov2}, and \texttt{cov3} use 13-bit, 14-bit, and 15-bit instrumentation widths, respectively, as evaluated in Fig.~\ref{Fig:cover_instrumentation}. All configurations have optimization enabled for higher coverage, as shown in Fig.~\ref{Fig:cov_instrument_fuzz}. Note that \texttt{cov3} is used in our final implementation, as reported in Table~\ref{tab:fpga_utilization}.

\begin{table}[H]
\centering
\caption{Resource Usages of Different Modules}
\label{tab:fpga_utilization}
\resizebox{\columnwidth}{!}{
\begin{tabular}{lccccc}
\toprule
\textbf{Resource} &
\makecell{\textbf{Rocket}\\\textbf{(DUT)}} &
\makecell{\textbf{Fuzzer\ IP}} &
\textbf{TurboFuzz} &
\makecell{\textbf{ILA}\\\textbf{(config1)}} &
\makecell{\textbf{ILA}\\\textbf{(config2)}} \\
\midrule
LUTs         & 308739  & 67523   & 89394     & 8142  & 10078 \\
             & (59.09\%) & (12.92\%) & (17.11\%)   & (1.56\%) & (1.93\%) \\
\midrule
Block RAMs   & 20      & 176     & 227     & 465   & 578 \\
             & (2.03\%)  & (17.89\%) & (23.07\%)   & (47.26\%) & (58.74\%) \\
\midrule
Registers    & 170400  & 91445   & 139477    & 14294 & 17322 \\
             & (16.30\%) & (8.75\%)  & (13.34\%)   & (1.37\%) & (1.66\%) \\
\bottomrule
\end{tabular}}

\end{table}

\section{Conclusion}

This paper introduces TurboFuzz—an end-to-end hardware-accelerated verification framework for processors. TurboFuzz implements the complete verification loop on a single FPGA, including coverage-driven stimulus generation, self-checking, and coverage collection, significantly improving verification efficiency. The framework also features several innovations: a hybrid fuzzing strategy, optimized corpus scheduling, and enhanced coverage instrumentation for better state reachability. Experiments on modern RISC-V processor cores demonstrate that TurboFuzz achieves $2.23\times$ greater coverage and $571\times$ performance improvement compared to state-of-the-art software simulation-based fuzzing methods, while also uncovering previously undetected bugs.

Future work includes developing resource-aware hardware coverage metrics for circuit state transitions and enhancing TurboFuzz's performance. We are working on porting TurboFuzz to verify larger RISC-V CPUs such as XiangShan\cite{xiangshan} based on disaggregated CPU+FPGA architecture or multi-FPGA platforms. We also plan to extend the framework's applicability to ISAs beyond RISC-V, particularly by generalizing the fuzzer module. Finally, we are committed to releasing TurboFuzz as open-source to support the broader hardware verification community.

\section*{Acknowledgment}
This work was supported by the National Key Research and Development Program of China (Grant No.2023YFB4405105) and the
Major Program of the National Natural Science Foundation of China
(Grant No.62090023).

\clearpage

\bibliographystyle{IEEEtranS}
\bibliography{refs}

\end{document}